\documentclass{appolb}
\usepackage{epsfig}
\usepackage{amsmath}
\usepackage{slashed}

\begin{document}
\title{Conundrums at Finite Density
\thanks{Presented at the international conference on ``Criticality in QCD and 
the Hadron Resonance Gas'', 29-31 July 2020, Wroclaw Poland.}
\headtitle{Conundrums at finite density} 
\headauthor{Rajiv V. Gavai}
}
\author{ Rajiv V. Gavai\thanks{Address after 1 January 2021 : Indian Institute
of Science Education \& Research, Bhopal bypass, Bhauri, Bhopal 462066, India.}
\address{Fakult\"at f\"ur Physik\\ Universit\"at Bielefeld\\
33615 Bielefeld\\ Germany }}
\maketitle
\begin{abstract}
Extending the successes of lattice quantum chromodynamics(QCD) at zero as well
as nonzero temperatures to nonzero density is extremely desirable in view of 
the quest for the QCD phase diagram both theoretically and experimentally. It
turns out though to give rise to some conundrums whose resolution may assist
progress in this exciting but difficult area, and should therefore be sought
actively.
\end{abstract}

\PACS{12.38.Gc, 11.35.Ha, 05.70.Jk}  
\preprint{TIFR/TH/16-38}

\section{Introduction}

The theory of strong interactions, Quantum Chromo Dynamics (QCD), has intriguing
properties such as  confinement or chiral symmetry breaking which have been
enigmas for over half a century.  A major reason is, of course, the dominance of
large coupling in such hadronic properties. Crucial clues in building physical
pictures to understand them were provided by the studies of simple models such
as the bag model or NJL-model. The discovery of instanton solutions and the
subsequent investigations of instanton-based models enhanced our understanding
further by emphasizing the role of the zero and near-zero modes of the Dirac
equation for interacting quarks.  Investigating all these models in extreme
environments such at high temperatures/densities led to a variety of phase
diagrams of strongly interacting matter.   It may not come as a surprise that
even qualitative features of these model phase diagrams differed substantially,
not to mention the quantitative details. For instance, the early sketches of the
QCD phase diagram display separate deconfinement and chiral transitions for all
temperatures and densities~\cite{baym}.  Nevertheless, they pointed to an
interesting path to fathom chiral symmetry breaking and/or confinement.

QCD formulated on a space-time lattice has yielded a more firm guidance in
refining these pictures at finite temperatures to give us a reliable, in some
cases even quantitative knowledge of the phase structure.  However, extending
this to finite densities or equivalently nonzero chemical potential, one
encounters conundrums many of which are unrelated to the latticization and were
hitherto still unknown to exist. These pose significant hurdles in excursions
inside the diagram from the temperature axis.  There is, of course, the famous
fermion sign(phase) problem at nonzero baryon density or equivalently nonzero
baryon chemical potential.  The aim of this talk is to draw attention to the
other, perhaps equally serious, problems.

\section{The $\mu \ne 0$ problems : I. Divergences }

\begin{figure}[htb]
\begin{center}
\includegraphics[scale=0.4]{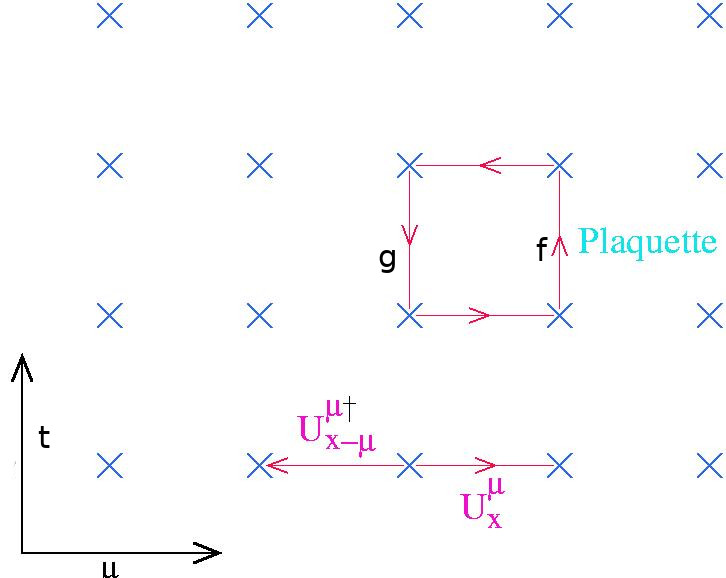}
\end{center}
\caption{Space-time(inverse temperature) lattice depicting a smallest loop
(plaquette) with time links and a quark covariant derivative term.}
\label{lat}
\end{figure}

\begin{figure}[htb]
\begin{center}
\includegraphics[scale=0.5]{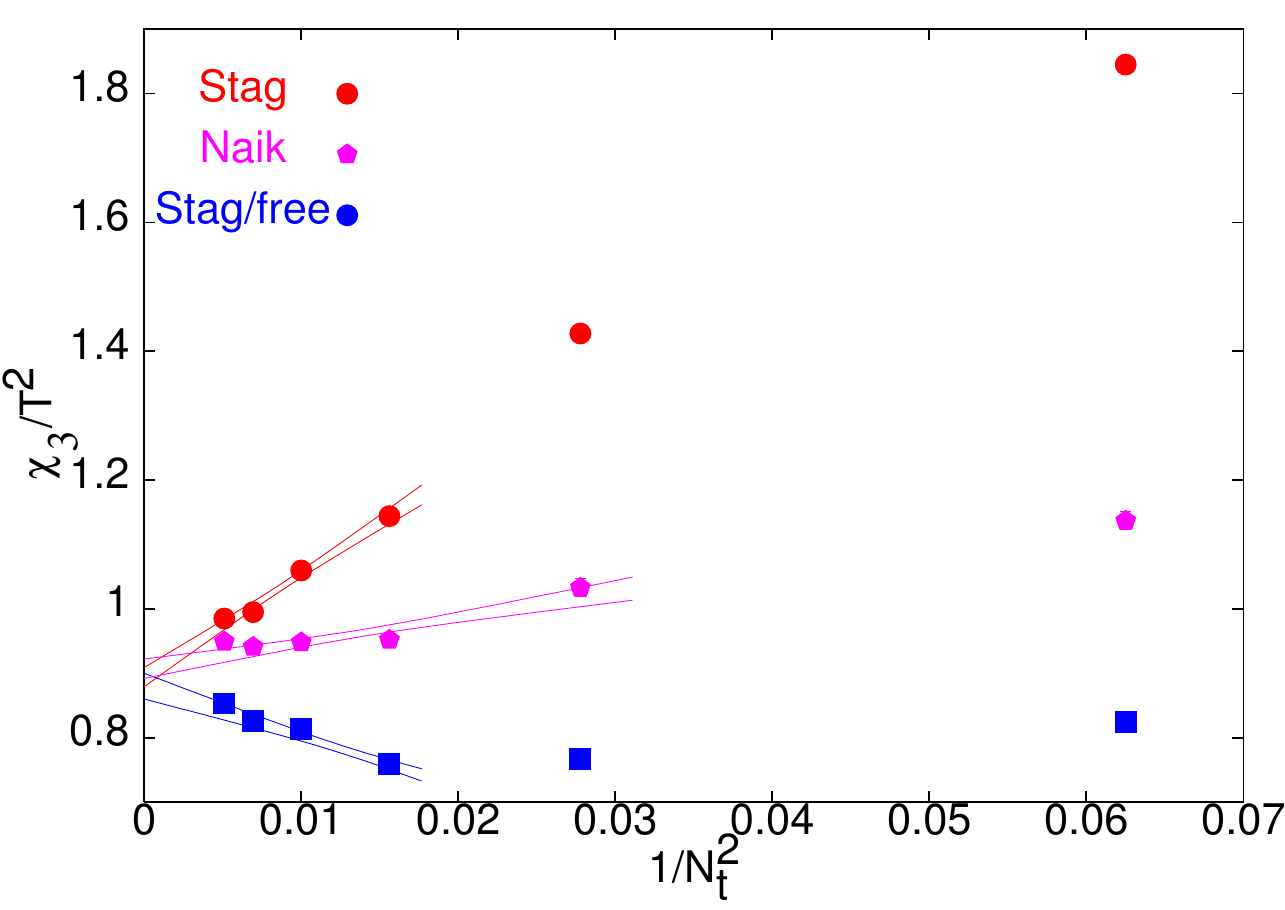}
\end{center}
\caption{Continuum limit for quark number susceptibilities with different
actions. A linear behaviour of the data and convergence to a unique continuum 
limit indicates the absence of any divergence. Taken from Ref.~\cite{GaGu}.}
\label{univ}
\end{figure}

Let us begin by recalling that the (baryonic) chemical potential is a Lagrange
multiplier to enforce the constraint of (net baryon) number conservation in the
grand canonical ensemble: $\partial_\mu J^B_\mu = 0$ is the current conservation
equation and $N_B = \int ~ d^3x ~J^B_0$ is the conserved charge. Following the
same principle on lattice, one obtains~\cite{BilGav} a point split version for
the conserved number.  Thus introducing the chemical potential on the lattice
amounts to multiplying the forward [backward] time-like links with $f(\mu a)$
[$g(\mu a)$], with $f(\mu a) = 1 + \mu a$ [$g(\mu a) = 1 - \mu a$] as seen in
Figure \ref{lat} for the na{\i}ve fermions. This form of $f$ and $g$, which has
been shown to remain the same for Wilson/Staggered/Improved local fermions as
well, leads~\cite{BilGav} to the following form of the energy density and quark
number density for a gas of free quarks:
\begin{eqnarray}
\epsilon &=& c_0a^{-4} + c_1\mu^2a^{-2} + c_3\mu^4 + c_4\mu^2T^2 + c_5T^4 \\
\nonumber n &=& d_0a^{-3} + d_1\mu a^{-2} + d_3\mu^3 + d_4\mu T^2 + d_5T^3. 
\end{eqnarray} 
Here $c_i$ and $d_i$ are constants, $a$ is the lattice spacing, and the
subscript B of $\mu$ has been dropped for simplicity as well as to indicate that
these expressions hold for any conserved charge such as strangeness or electric
charge.  In the continuum limit of $ a \to 0$, one obtains a leading quartic
divergence and a subleading $\mu$-dependent quadratic one.  Subtracting off the
vacuum contribution at $T = 0 = \mu$ eliminates the leading divergence in each
case.  However, the $\mu$-dependent $a^{-2}$ divergences persist in both the
energy density and the quark number density.  Note that these divergences are
present for the free theory itself.  As a solution to this problem different
forms of $f$ and $g$ have been proposed.  The popular exponential
choice~\cite{HaKaKo}, $f(\mu a) = \exp(\mu a)$ and $g(\mu a) = \exp(-\mu a)$ as
well as another choice~\cite{BilGav}, $f(\mu a) = (1 + \mu a)/ \sqrt(1- \mu^2
a^2)$ along with $g(\mu a) = (1 - \mu a)/\sqrt(1- \mu^2 a^2)$, both lead to
their corresponding $c_1 = 0 = d_1$, which then lead to finite results in the
continuum limit.  Indeed the $\mu$-dependent divergences are eliminated for all
$ f \cdot g = 1$~\cite{Gav85}.  One anticipates this analytical proof of the
lack of $\mu$-dependent divergences for free quarks to hold true in any
order-by-order perturbative inclusion of interactions with gluons. However,
numerical simulations are needed, and were employed~\cite{GaGu}, to extend the
proof for the non-perturbative interacting case as well, as shown in Figure
\ref{univ}. Both the lack of any diverging behaviour as well as a unique
continuum limit is evident in the data.

A natural question arises as to why are there three (or more) lattice QCD
actions when the continuum QCD has only one.  The usual answer is universality.
As long as all these actions reduce to the continuum QCD action in the limit of
$a \to 0$, universality tells us that physics should be the same for all of them
in that limit. Expanding the functions $f$,$~g$ in powers of $\mu a$, one finds 
that the three actions differ by terms of ${\cal O} (\mu^2a^2)$  and higher 
which vanish in the $a \to 0$ limit and are thus {\em irrelevant} terms.

{\bf Paradox :} These irrelevant terms vanish from action as $ a \to 0$ but 
do eliminate divergences. This appears to be a violation of universality!
On the other hand, since the divergence cancelling terms are absent in the
continuum theory, as for the na{\i}ve case, one wonders whether the divergences
are present in the continuum theory itself.  As for any usual improved action,
one hopes that universality will ensure physical results are unaffected but it
seems prudent to check it in view of the above conundrum.

These modified/improved actions have a further problem. One can work out the
current conservation equation for the Lagrangian with $\mu \ne 0$.  It remains
unchanged only for the linear $\mu$-case.  It acquires $\mu^2 a^2$  and higher
order terms of even powers in all the other cases.  Thus integrating over
spatial dimensions, one obtains a conserved charge on the lattice only for the
linear $\mu$-form.  For all the divergence eliminating form of actions one has
{\bf no} conserved charge on the lattice anymore!  Consequently, $ {\cal Z} \ne
\exp (-\beta[\hat H-\mu \hat N])$ on the lattice for them, and therefore one
cannot define an {\em exact} canonical partition function on lattice from the
${\cal Z}$ defined this way. ${\cal Z} = \sum_n z^n {\cal Z}^C_n $ on the
lattice only for the na{\i}ve linear $\mu$-action.  Once again, one has to hope
that it is possible at least in the continuum limit of $a \to 0$ but clearly an
explicit demonstration is necessary.

Most computational methods, if not all, consist of integrating out the quark
fields, leading to the quark determinant.  Due to its gauge-invariant nature, 
the determinant can be seen as a sum over all possible quark loops.   
Any $\mu$-dependence for ${\cal Z}$ arises solely due to loops with 
time-like links, and hence is $\propto (f \cdot g)^l $, where $l$ is the number
of positive timelike links in the loop. This is illustrated for the simplest
case of $l=1$ in Figure \ref{lat}.  Quark loops of {\em all} sizes and types 
contribute for the na{\i}ve case of $f$,$~g$ $= 1 \pm \mu a$, as is indeed 
the case also in the continuum. However, since $f \cdot g = 1 $ for the other
two actions, it is clear that only limited number of loops contribute.  Indeed,
only quark loops winding around the $T$-direction contribute to $\mu_B$
dependence for these cases. Again if all the actions were to lead to the {\em
same} physics, as they ought to, small quark loops which are topologically
trivial must start also contributing, as $a \to 0$.  It is far from clear how
this may happen since for all non-vanishing $a$, the  $f \cdot g = 1 $ condition
applies and these loops do {\em not} contribute to any $\mu$-dependence.  One
possible way out maybe that the small loops sum up to a $\mu$-independent
constant, preferably zero.  It is far from clear how this might come about in
the interacting theory.  This is yet another conundrum which universality
suggests should resolve itself in the continuum limit, and needs to be verified
by explicit computations.

\section {Divergences exist in the continuum too}

Recall that the conundrums discussed in the section above were related to the
differences in the $f(\mu)$ and $g(\mu)$ : $f \cdot g = 1 - \mu^2 a^2$ for the
na{i}ve linear case and $f \cdot g = 1$ for the other two.  This, in turn,
arose, as the latter got rid of the $\mu$-dependent divergences that arise for
the former choice.  Since in the continuum limit one finally has the only linear
form, one may wonder whether the $\mu$-dependent divergences exist in the
continuum as well, and the lattice as a regulator is merely reproducing them
systematically or whether the latticization itself introduces the divergences.

Indeed, it turns out that contrary to the common belief, the free theory
divergences are {\bf not} lattice artifacts. They exist in continuum too.
Instead of the lattice regulator, one can employ a momentum cut-off $\Lambda$ in
the continuum theory to show~\cite{GaSh15} the presence of $\mu \Lambda^2$ terms
in number density easily. We summarise below why one ought to expect them in
the continuum itself.

\begin{figure}[htb]
\includegraphics[scale=0.6]{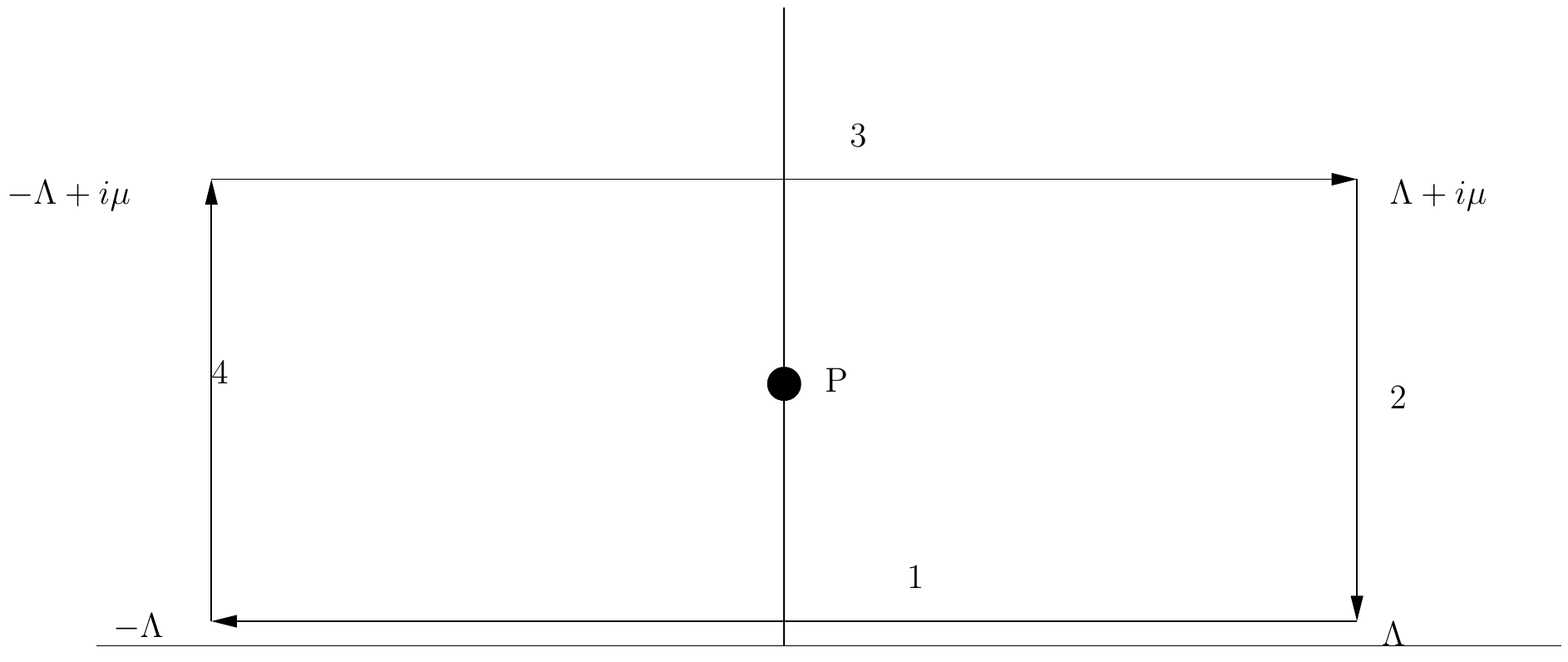}
\caption{The contour diagram for calculating the number density for free 
fermions at zero temperature. P denotes the pole.}
\label{cntu}
\end{figure}

The quark number density, or equivalently  third of the baryon 
number density for a single flavour, is defined as,
\begin{equation}
 n= \frac{T}{V}\frac{\partial \ln \mathcal{Z}}{\partial \mu}|_{T={\rm fixed}}
\end{equation}
with $\mathcal{Z}$ for free fermions given by
\begin{equation}
 \mathcal{Z}=\int \mathcal{D}\bar \psi \mathcal{D} \psi
\rm{e}^{\int_0^{1/T} d\tau \int d^3 x\left[-\bar\psi
(\gamma_\mu \partial_\mu + m -\mu \gamma_4)\psi\right]},
\end{equation} 
Evaluating the quark number density, $n$ in the momentum space for the
massless free quark gas, one has
\begin{equation}
\nonumber
 n=\frac{2iT}{ V}\sum_{m}\int \frac{d^3p}{(2\pi)^3}\frac{(\omega_m-i\mu )}
 {p^2+(\omega_m-i\mu )^2}~ \equiv \frac{2iT}{ V}\int \frac{d^3p}{(2\pi)^3} \sum_{\omega_m}
 F(\omega_m,\mu ,\vec p),
\nonumber
\end{equation}
where $p^2= p_1^2+ p_2^2+ p_3^2$  and $\omega_m=(2m+1)\pi T$.  
In the usual contour method, the sum over $m$ or $\omega_m$ gets replaced as an 
integral in the complex $\omega$-plane.  Together with the subtracted 
vacuum ($\mu$=0) contribution, one has in the complex $\omega$-plane line
integrals along the directed arms 3 and 1 in Figure~\ref{cntu}. Adding and
subtracting the side arm line integrals, one obtains the canonical answer from
the residue of the pole $P$ in Figure~\ref{cntu}.  One still has to evaluate the
side arm contributions.

Introduce a cut-off $\Lambda$ for all 4-momenta at $T=0$ for a careful
evaluation of the divergent arms 2 \& 4 contributions in the Figure~\ref{cntu}.
The $\mu \Lambda^2$ terms can be seen to arise~\cite{GaSh15} from the arms 2 
\& 4.  

\begin{equation}
 \int\frac{d^3p}{(2\pi)^3}
\Big( \int_2+\int_4 \Big) \frac{d \omega}{\pi} \frac{\omega} {p^2 +\omega^2}
= - \frac{1}{2\pi} \int \frac{d^3p}{2\pi^3} \ln \left[ \frac{p^2+
(\Lambda+i \mu)^2}{p^2+(\Lambda-i \mu)^2}\right].
\end{equation}

Utilising the fact that $\Lambda \gg \mu$, the integrand can be expanded in
$\mu/\Lambda$ to discover that while the leading $\Lambda^3$ terms do indeed
cancel there is a nonzero coefficient for the subleading $\mu \Lambda^2$ term.
It may be worth noting that the arms 2 \& 4 make a finite contributions to the
$\mu^3$ term as well. One usually ignores the subleading contribution from the
arms 2 \& 4, amounting to a subtraction of the `free theory divergence' in the
continuum. This practice suggests a prescription of subtracting the free theory
divergence by hand on the lattice as well.  Such a prescription surely works in
including the interactions in a perturbation theory.  It has been tested in
numerical simulations, and found to work excellently.

In order to test whether the divergence is truly absent in simulations, one
needs to take the continuum limit $a \to 0$ or equivalently $N_T \to \infty$ at
fixed $T^{-1} = a N_T$.  For quenched QCD at $T/T_c= 1.25$ \& 2  and for quark
mass $m/T_c =0.1$, lattices  with $N_T = 4$, 6, 8, 10 and 12 were
employed~\cite{GaSh15}. On 50-100 independent configurations quark number
susceptibility was computed.  Since it is a derivative of the number density
with $\mu$, it should have a simple $a^{-2}$ divergence.  The $1/a^2$-term for
free fermions on the corresponding $N^3 \times \infty$ lattice was subtracted
from the computed values of the susceptibility in simulations.  The results are
displayed in Figure \ref{qns} as a function of $1/N_T$.  If the interactions
were to induce additional non-perturbative divergent contribution over and above
the subtracted free theory ones, the susceptibility should behave as
$\chi_{20}/T^2= c_2(T) N_T^2 + c_1(T) + c_3(T) N_T^{-2} + \mathcal{O}(N_
T^{-4})$. The divergent $c_2$-contribution would then lead to a rapid shoot-up
near the $\chi_{20}/T^2$-axis. $c_1$ is the expected continuum result with $c_3$
governing the approach to the limit.

\begin{figure}[htb]
\includegraphics[scale=0.5]{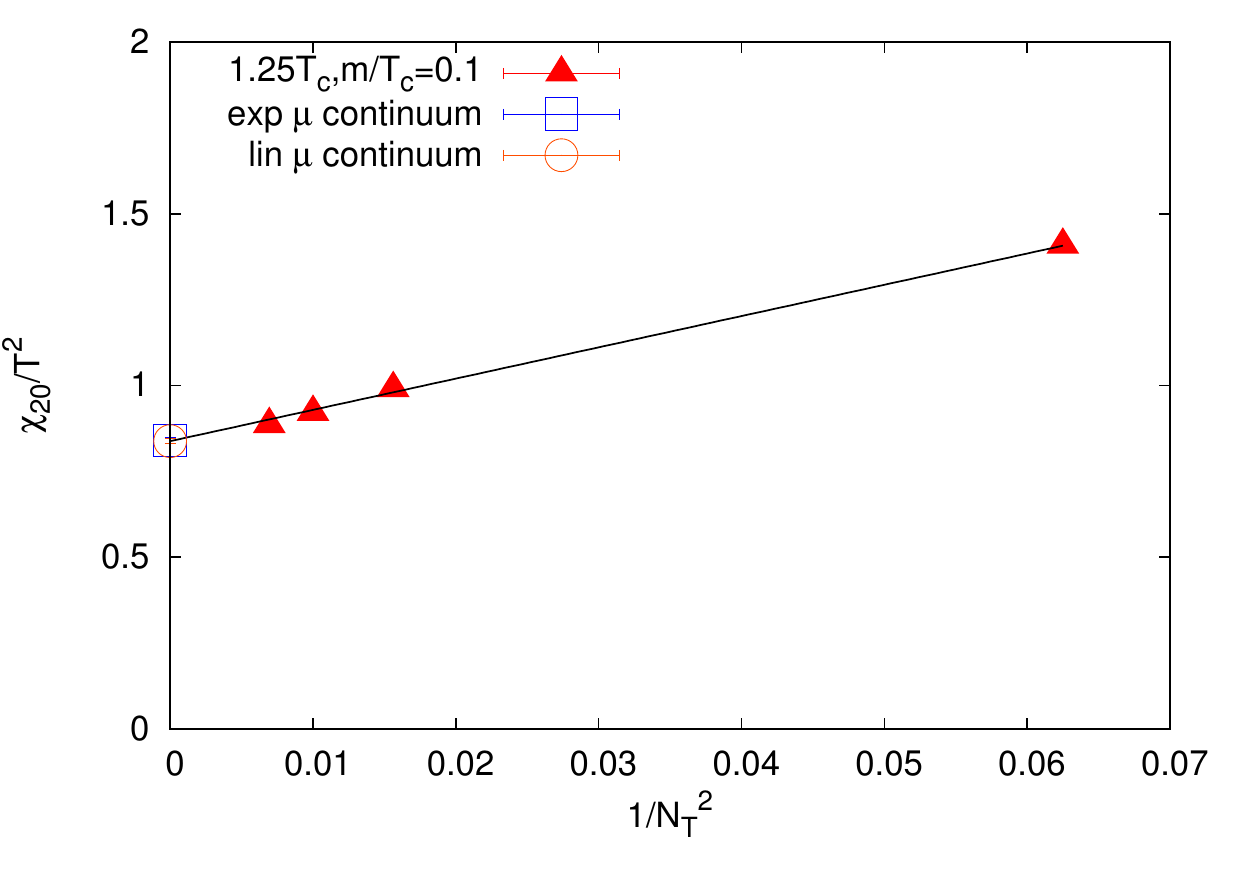}
\includegraphics[scale=0.5]{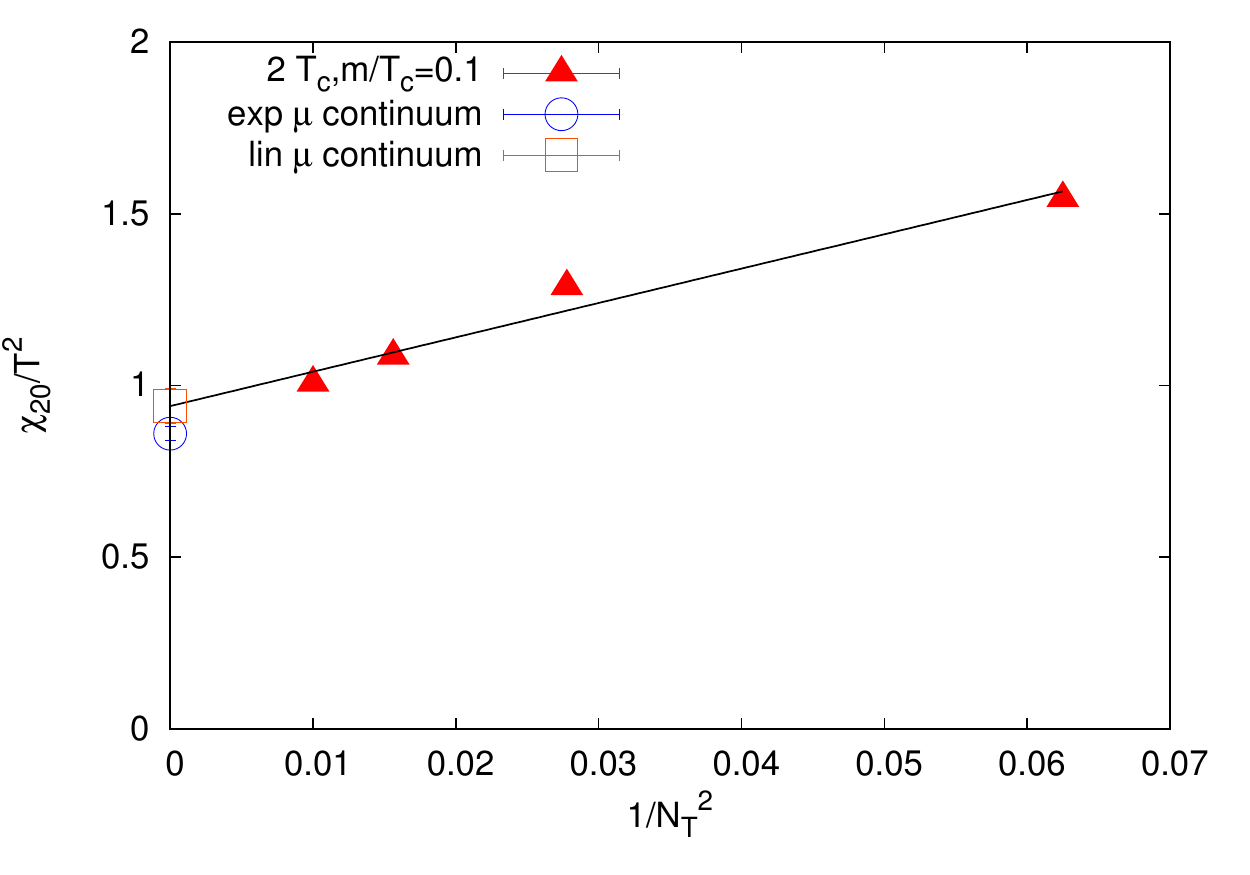}
\label{qns}
\caption{The quark number susceptibility at $1.25T_c$ (left panel) and $2T_c$ 
(right panel) for $m/T_c=0.1$. Taken from Ref.~\cite{GaSh15}}
\label{qns}
\end{figure} 

A glance at both the panels of the Figure \ref{qns} shows an evident lack of any
divergent rise in both or equivalently $c_2 \simeq 0$ for both temperatures,
since both sets of data display only positive slope throughout.  Furthermore,
the extrapolated continuum result coincides with the earlier result obtained
with the $\exp(\pm a \mu)$ action~\cite{sm06}.

\section{The $\mu \ne 0$ problem : II. Quark Type}

Placing quark fields on a lattice has the famous doubling problem.  Mostly
staggered quarks are used in lattice QCD simulations, as they possess some
chiral symmetry.  Consequently, the chiral condensate, $\langle \bar \psi \psi
\rangle$ can be employed as an order parameter to investigate the QCD phase
diagram as a function of $T$ and $\mu_B$.  However, flavour and spin symmetry
are broken for them.  Moreover, flavour singlet $U_A(1)$ symmetry is broken
explicitly and thus the question of the $U_A(1)$ anomaly is mute.   On the other
hand, the holy grail of phase diagram, namely, the QCD critical point needs two
light flavours and the anomaly to persist~\cite{PiWi} for the chiral transition
on the $\mu_B =0$ axis to be of second order, and hence it to be a cross over
for physical light quarks. Domain Wall or Overlap quarks are therefore a better
choice due to their ``exact'' chiral symmetry on the lattice. Although their
nonlocality makes them computationally expensive, one can at least in principle
employ them to study the QCD critical point.  Defining chemical potential for
them turns out, however, to be tricky.  In particular, introducing chemical
potential for either of them faces yet another conundrum related to the
divergence problem discussed above.   

The usual Wilson Dirac fermion matrix is at the heart of definition of both
these nonlocal quarks.  Adapting the exponential prescription for $D^{\rm
Wilson}$, Bloch and Wettig~\cite{BlWe} introduced $\mu$.  This definition was
shown to have no divergences in the free theory~\cite{GaLi,BGS}. Unfortunately
the BW-prescription breaks the lattice chiral symmetry at any finite
density~\cite{BGS}, leaving us without any order parameter.

Luckily, a lattice action with continuum-like chiral symmetries for quarks at
nonzero $\mu$  has been proposed already~\cite{GaSh12}.  Since the massless
continuum QCD action for nonzero $\mu$ can be written explicitly as a sum over
right and left chiral modes of quarks, the key idea was to employ similar chiral
projections for the Overlap quarks to construct the action at nonzero $\mu$.  It
was shown to have exact chiral invariance on the lattice, and thus chiral
condensate works as an order parameter for the entire $T$-$\mu_B$ plane
~\cite{GaSh12}.  Moreover, using the Domain Wall formalism, it was also shown
why this is physically the right thing to do: it counts only the physical (wall)
modes as the baryon number while the BW action includes all the unphysical heavy
modes as well.

It turns out, however, that this chirally invariant Overlap action with nonzero
$\mu$ is linear in $\mu$, i.e., comes with divergences.  Furthermore, inventing
the $f$,$~g$ in this case which will i) eliminate the divergences and ii) still
preserve the exact chiral invariance on the lattice has so far not been
possible.  Recently, it has been shown that SLAC fermions, which too are
nonlocal but possess exact chiral symmetry, also need a linear form in $\mu$ at
finite density, and it too possesses these divergences~\cite{Le20}.  Thus the
linear form seems the natural physical choice if chiral symmetries are to be
exact on the lattice, although the resultant free theory has divergences.  As in
the previous section, these divergences can always be subtracted out especially
if eliminating them using nonlinear forms for $f$, $g$ leads to the conundrums
already discussed.

\section{The $\mu \ne 0$ problems : III. Topology} 

Instanton vacuum provides a nice physical picture of chiral symmetry breaking
and the chiral phase transition~\cite{ssd}.  Overlap Dirac operator spectra
have been used to investigate topology and to understand the nature of the high
temperature phase.  In particular,  the number of low quark eigenmodes get
depleted ~\cite{ZeTQ} as $T$ goes up and the number of exact zero modes, a
measure of topological susceptibility, falls exponentially in the quark-gluon
plasma phase. Naturally, one can envisage doing a similar study for the high 
density phase.  However, it is not easy for QCD due to the sign problem.

QCD at nonzero isospin density as well as two colour QCD do not have a sign
problem, as the quark determinant is real in each of these cases.  A lot of work
on both the cases has been done in studying the phase
structure~\cite{MuI,MuNc2}.  In both these theories, it has been observed that
an increase in number density and a drop in the chiral condensate occurs 
at the same $\mu_c$.  Interestingly though the spectra of the corresponding low
quark modes appear unaffected~\cite{MuI1,MuNc21} as a function of the 
corresponding $\mu$ even when one runs through $\mu_c$ restoring the chiral 
symmetry.  This observation in two different theories 
raises an interesting  possibility that the chiral symmetry 
restoration is decoupled from a change in topology, and thus from perhaps the 
deconfinement transition, at finite density/chemical potential in general.

Figure~\ref{zemui} displays the eigenvalue distribution~\cite{MuI1} on the 
log scale to highlight differences in the near-zero modes in the low 
and high density phases for the nonzero isospin case:

\begin{figure}[htb]
\includegraphics[scale=0.60]{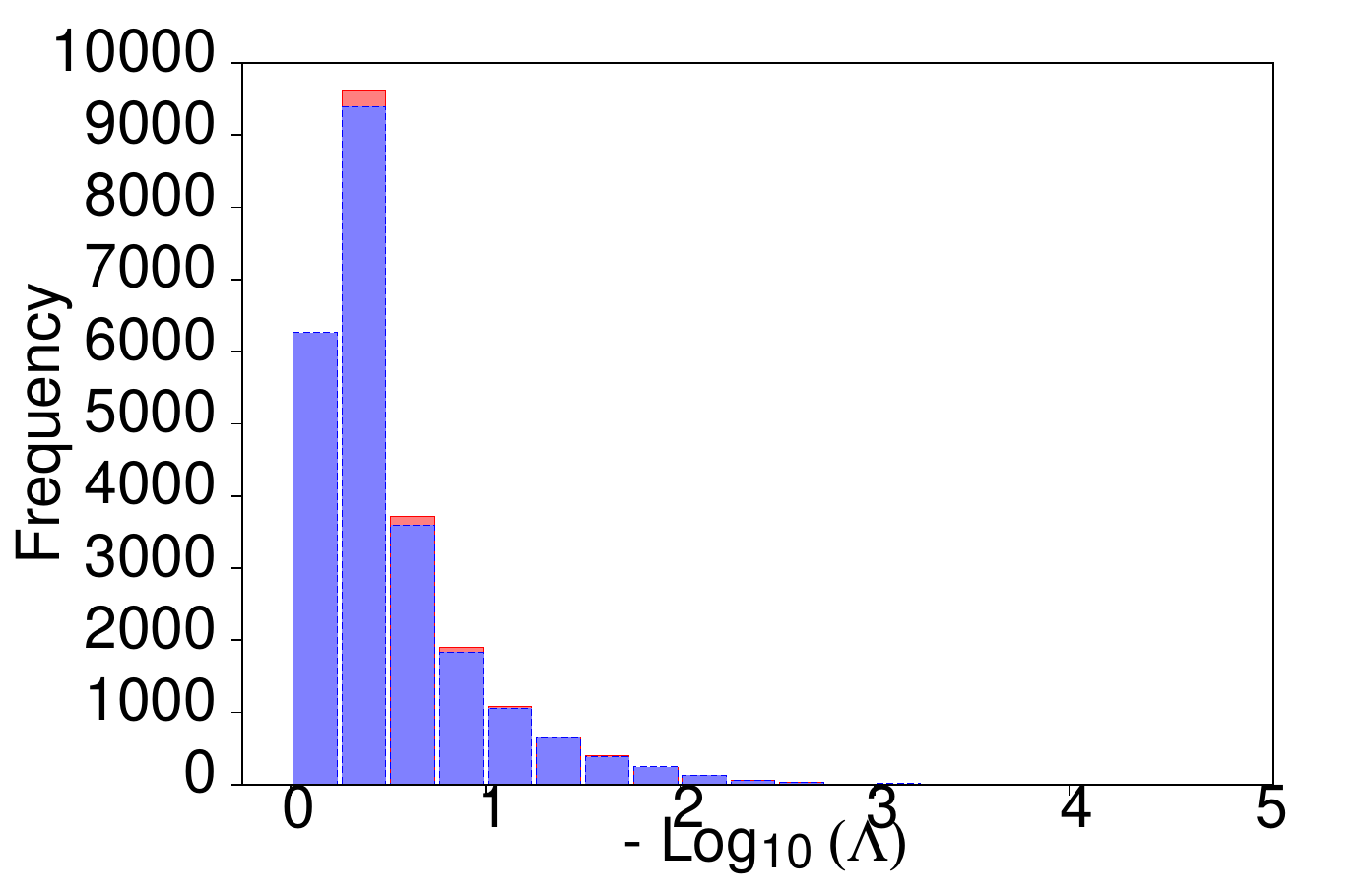}
\caption{A comparison of the near-zero quark mode distributions below (blue) and
above (red) the finite isospin chemical potential at which chiral symmetry
restoration occurs.  No visible difference is evident. Taken from
Ref.~\cite{MuI1}, where further details can be found.}
\label{zemui}
 \end{figure}

Similarly very little or no change is visible in the number of exact zero modes,
or equivalently, the topological susceptibility in both the
cases~\cite{MuI1,MuNc21} across the corresponding chiral symmetry restoring
transition.

\section{Summary} 

Investigations at finite density using the reliable lattice QCD techniques face
many hurdles, the most famous of which is the sign/phase problem of the quark
determinant. We pointed out that the introduction of the chemical potential on
the lattice itself is plagued with conundrums.  Most of these, including the
$\mu$-dependent divergence, are not due to latticization. Indeed, lattice only
reproduces faithfully what exists in the continuum field theory.  Elimination of
the divergence by modifications of action, as is commonly done, leads to
apparent conflicts with universality which need to be resolved by carrying out
continuum limit computations for many different ways of adding chemical
potential.  

Chiral and flavour invariance is crucial for the QCD critical point
investigations.  Eventually one will have to employ the overlap quarks at finite
density for reliable simulations.  Doing so while retaining the chiral symmetry
seems to lead to a linear $\mu$-dependent action always.  Subtraction of free
theory divergences was demonstrated to suffice nonperturbatively and should be
tested for the overlap action as well.

Numerical simulations suggest that the distribution of the topological charge,
$Q$, changes very little in going from the low $T$ \& low density phase to the
low $T$ \& high density phase as one goes across the isospin chemical potential
$\mu_I$ or $\mu^{N_c=2}$ phase transitions, although the chiral condensate drops
and number density picks up at each of these phase transition.  This is  in
contrast to the change of low $T$ to high $T$ phase, which exhibits an
(exponential) fall-off. This may be a hint towards a possible separation of the
chiral symmetry restoring transition and the deconfining phase at finite
density.  It will be challenging to check if this is indeed so for the finite 
density QCD.

\section{Acknowledgements}

It is a pleasure to gratefully acknowledge the financial support by the 
Deutsche Forschungsgemeinschaft (DFG) through the 
the Project grant No.  315477589-TRR  211 (Strong-interaction matter under 
extreme conditions).

\end{document}